\documentstyle[aps,preprint,eqsecnum]{revtex}

\begin{document}
\draft
\title{Entangled SU(2) and SU(1,1) coherent states}
\author{Xiao-Guang Wang$^1$, Barry C.\ Sanders\thanks{%
Email: barry.sanders@mq.edu.au}$^2$, and Shao-hua Pan$^1$}
\address{$^1$Chinese Center of Advanced Science and Technology (World Laboratory),\\
P.O.Box 8730, Beijing 100080 \\
and Laboratory of Optical Physics, Institute of Physics, \\
Chinese Academy of Sciences, Beijing 100080, P.\ R.\ China \\
$^2$Department of Physics, Macquarie University, Sydney, New South Wales 
2109, Australia}
\date{\today}
\maketitle

\begin{abstract}
Entangled SU(2) and SU(1,1) coherent states are developed as superpositions
of multiparticle SU(2) and SU(1,1) coherent states. In certain cases, these
are coherent states with respect to generalized su(2) and su(1,1)
generators, and multiparticle parity states arise as a special case. As a
special example of entangled SU(2) coherent states, entangled binomial
states are introduced and these entangled binomial states enable the
contraction from entangled SU(2) coherent states to entangled harmonic
oscillator coherent states. Entangled SU(2) coherent states are discussed in
the context of pairs of qubits. We also introduce the entangled negative
binomial states and entangled squeezed states as examples of entangled
SU(1,1) coherent states. A method for generating the entangled SU(2) and
SU(1,1) coherent states is discussed and degrees of entanglement calculated.
Two types of SU(1,1) coherent states are discussed in each case: Perelomov
coherent states and Barut-Girardello coherent states.
\end{abstract}



\section{Introduction}

\label{sec:intro}

Qubits are the basic elements of quantum information technology, in the same
way that bits are basic units of information in computers. Whereas bits are
binary digits, qubits are spin-$1/2$, or two-level, quantum systems. The
advantage of quantum computing over classical computing is the capacity for
producing entangled qubits: the large state space available for entangled
qubits enables certain problems, thought not to be computable on classical
computers, to be solved on quantum computers\cite{Steane98}.

A bit can be in an off, or `0', state or an on, or `1', state, but the qubit
can be in a superposition of an off, or `$|0\rangle $', state and an on, or `%
$|1\rangle $', state. We can represent such a state (a general qubit) by 
\begin{eqnarray}
|\theta ,\phi \rangle &=&\exp \left[ -\frac \theta 2(\sigma _{+}e^{-i\phi
}-\sigma _{-}e^{i\phi })\right] |1\rangle  \nonumber \\
&=&\cos \frac \theta 2|1\rangle +e^{i\phi }\sin \frac \theta 2|0\rangle ,
\end{eqnarray}
up to a global phase. Here $\sigma _{\pm }=\sigma _x\pm i\sigma _y$ for $%
\sigma _x$ and $\sigma _y$ are Pauli matrices. In fact such a state is an
SU(2) coherent state, also known as an atomic coherent state \cite
{Are72,Per86}. Two qubits, prepared in a product state, could then be
expressed as $|\theta _1,\phi _1\rangle \otimes |\theta _2,\phi _2\rangle $.
However, quantum computation is based on the exploitation of entanglement,
and such product states are of limited value. The simplest extension of this
arbitrary two-qubit product state to a two-qubit entangled state is the
unnormalized state 
\begin{eqnarray}
&&\cos \theta |\theta _1,\phi _1\rangle \otimes |\theta _2,\phi _2\rangle 
\nonumber \\
&&+e^{i\phi }\sin \theta |\theta _1^{\prime },\phi _1^{\prime }\rangle
\otimes |\theta _2^{\prime },\phi _2^{\prime }\rangle ,
\end{eqnarray}
which is a product state for $\theta $ a multiple of $\pi /2.$ Of course the
Bell states\cite{Bell65,Bra92} 
\begin{eqnarray}
|\Phi \rangle _{\pm } &=&\frac 1{\sqrt{2}}\left( |0\rangle \otimes |0\rangle
\pm |1\rangle \otimes |1\rangle \right) ,  \nonumber \\
|\Psi \rangle _{\pm } &=&\frac 1{\sqrt{2}}\left( |0\rangle \otimes |1\rangle
\pm |1\rangle \otimes |0\rangle \right)
\end{eqnarray}
are special cases of the general state (1.2).

The SU(2) coherent states form an overcomplete basis for the Hilbert space,
and the qubit states correspond to spin-$1/2$ representations of SU(2).
There are therefore subtleties concerning this entanglement of
non-orthogonal states: such subtleties have been considered with respect to
entangled coherent states (or superpositions of multimode coherent states)
where the coherent states have been harmonic oscillator coherent states\cite
{Tom87,San92,Cha92,Wie93,Ans94,Man95,San95,Jex95,Spi95,Ger97,Rai97,Guo97,Bos97,Ric98,San99,Kim99,Gil99}%
. Our objective here is to introduce and analyze entangled SU(2) and SU(1,1)
coherent states.

The entangled SU(1,1) coherent states are closely related to the SU(2)
coherent states because the algebra su(1,1) and su(2) are so similar.
However, there are two commonly considered coherent states for SU(1,1). One
SU(1,1) coherent state is the analog of the harmonic oscillator coherent
state achieved by displacing the vacuum state and the SU(2) coherent state
obtained by ``rotating'' the lowest- or highest-weight state. The analogous
SU(1,1) coherent state is obtained via an SU(1,1) transformation of
lowest-weight state. This SU(1,1) coherent state is a member of Perelomov's
category of generalized coherent state, and we refer to this state as a
Perelomov SU(1,1) coherent state\cite{Per86}. The second SU(1,1) coherent
state, introduced by Barut and Girardello\cite{BGCS}, is the analog of the
harmonic oscillator coherent state being an eigenstate of the annihilation
operator; an SU(2) coherent state of this type does not exist due to the
SU(2) Hilbert space being finite. We treat entangled SU(1,1) coherent states
of both the Perelomov and Barut-Girardello types. As a special case of the
entangled Perelomov SU(1,1) coherent state, we obtain superpositions of
squeezed vacuum states\cite{Sanders89} and entangled squeezed states\cite
{San95}. Squeezed states are significant in quantum limited measurements,
quantum communications and exotic spectroscopy of atoms\cite{Loudon87}.

In association with the parity operator, this paper first develops parity
coherent states and entangled SU(2) and SU(1,1) coherent states in Section
II. Section III generalizes the parity coherent states and considers
nonlinear SU(2) and SU(1,1) coherent states. In Section IV, we discuss how
to represent entangled SU(2) and SU(1,1) coherent states in Fock space, and
we have obtained the entangled binomial states, entangled negative binomial
states and entangled squeezed states as well as the contraction of entangled
SU(2) and SU(1,1) coherent states to entangled harmonic oscillator coherent
states. Section V investigates how to generate the entangled coherent states
in Hamiltonian systems. Section VI discusses the degree of entanglement for
these entangled coherent states, and a conclusion is given in Section VII.
The Appendix gives the most general entangled SU(2) and SU(1,1) coherent
states with certain special cases, including the quantum Fourier transform
state of a product multiparticle SU(2) coherent state acquired by Shor's
algorithm\cite{Shor}.

\section{Entangled SU(2) and SU(1,1) coherent states}

\subsection{Entangled coherent states of the harmonic oscillator}

For the harmonic oscillator, a general unnormalized two-particle entangled
coherent state can be expressed as\cite{Wie93} 
\begin{eqnarray}
&&\cos \theta |\alpha _1\rangle \otimes |\alpha _2\rangle +e^{i\phi }\sin
\theta |\alpha _1^{\prime }\rangle \otimes |\alpha _2^{\prime }\rangle  
\nonumber \\
&&=[\cos \theta \exp (\vec{\alpha}\cdot \vec{a}^{\dagger }-\vec{\alpha}%
^{*}\cdot \vec{a})  \nonumber \\
&&+e^{i\phi }\sin \theta \exp (\vec{\alpha}^{\prime }\cdot \vec{a}^{\dagger
}-\vec{\alpha}^{\prime *}\cdot \vec{a})]|0,0\rangle ,
\end{eqnarray}
for $\vec{a}\equiv (a_1,a_2)$, $\vec{\alpha}\equiv (\alpha _1,\alpha _2)$
and~$|0,0\rangle =|0\rangle _1\otimes |0\rangle _2$ the vacuum state. It is
particularly helpful to concentrate on the balanced entangled coherent state 
\begin{equation}
2^{-1/2}\left[ |\alpha _1\rangle \otimes |\alpha _2\rangle +e^{i\phi
}|-\alpha _1\rangle \otimes |-\alpha _2\rangle \right] ,
\end{equation}
\newline
where the word ``balanced'' refers to each component in the superposition
having coefficients of the same magnitude. For $\phi =\pi /2$ such a state
can, in principle, be generated via a nonlinear interferometer\cite
{San92,San99} and is an eigenstate of the canonically transformed pair
annihilation operator\cite{Spi95} 
\begin{equation}
\vec{\tilde{a}}\equiv \Pi\vec{a}
\end{equation}
with 
\begin{equation}
\Pi=e^{i\pi \vec{a}^{\dagger }\cdot \vec{a}},\text{ }\Pi\vec{a}\Pi=-\vec{a}.
\end{equation}
Thus, the entangled coherent state (2.1) is actually a pair coherent state%
\cite{Aga88} with respect to the canonically transformed vector annihilation
operator\cite{Spi95}, although this pair coherent state does not have the
restriction that the number difference of the two modes is fixed.

The most general superposition of harmonic oscillator coherent states is\cite{San99}
\begin{equation}
\int \frac{d^{2N}\vec{\alpha}}{\pi ^N}f(\vec{\alpha})|\vec{\alpha}\rangle
\end{equation}
with $|\vec{\alpha}\rangle =|\alpha _1\rangle \otimes |\alpha _2\rangle
\otimes \cdot \cdot \cdot \otimes |\alpha _N\rangle $ the $N-$particle
coherent state. The superposition (2.5) is entangled when it cannot be
expressed as a product state in any representation.

We can introduce an entangled SU(2) coherent state as a generalization of
(1.2). However, we make this analysis more general than necessary for
studying qubits. We wish to treat general irreducible representations~$j$,
where~$j$ is the angular momentum parameter and can be integer or half-odd
integer. The single qubit case corresponds to~$j=1/2$. A pair of qubits can
of course be treated as a single system: in this case we have one $j=0$
state (the singlet state) and $j=1$ states (the triplet states). The four
states together constitute the Bell states (1.3).

\subsection{Entangled SU(2) coherent states}

The SU(2) coherent state can be expressed as\cite{Are72,Per86,San89} 
\begin{eqnarray}
|j\text{ }\gamma \rangle &\equiv &R(\gamma )|jj\rangle  \nonumber \\
&=&\exp \left[ -\frac 12\theta \left( J_{+}e^{-i\phi }-J_{-}e^{i\phi
}\right) \right] |jj\rangle  \nonumber \\
&=&\left( 1+|\gamma |^2\right) ^{-j}\sum_{m=0}^{2j}{%
{2j \choose m}%
}^{1/2}\gamma ^m|j\,j-m\rangle ,
\end{eqnarray}
where $\gamma =\exp (i\phi )\tan (\theta /2),$ $R(\gamma )$ is the rotation
operator, and $J_{-}$ and $J_{+}$ are lowering and raising operators of the
su(2) Lie algebra, respectively. The su(2) generators $J_{\pm }$ and $J_z$
satisfy the su(2) commutation relations

\begin{equation}
\lbrack J_{+},J_{-}]=2J_z,\text{ }[J_z,J_{\pm }]=\pm J_{\pm }.
\end{equation}
We can define an su(2) parity operator as

\begin{equation}
\Pi =(-1)^{{\cal M}},\text{ }\Pi ^2=1,\text{ }\Pi ^{\dagger }=\Pi .
\end{equation}
Here ${\cal M}=J_z+j$ is the `number' operator such that

\begin{equation}
{\cal M}|j,m\rangle =(j+m)|j,m\rangle ,\text{ }0\leq j+m\leq 2j.
\end{equation}

It is easy to see that

\begin{equation}
\Pi J_{\pm }\Pi =-J_{\pm },\text{ }\Pi J_z\Pi =J_z.
\end{equation}
Using the above equation we have a new su(2) representation:

\begin{equation}
\lbrack \tilde{J}_{+},\tilde{J}_{-}]=2\tilde{J}_z,\text{ }[\tilde{J}_z,%
\tilde{J}_{\pm }]=\pm \tilde{J}_{\pm },
\end{equation}
where

\begin{equation}
\tilde{J}_{+}=J_{+}\Pi ,\text{ }\tilde{J}_{-}=\Pi J_{-}\text{, }\tilde{J}%
_z\equiv J_z.
\end{equation}
We define a new SU(2) coherent state associated with the su(2) algebra
(2.11) as

\begin{eqnarray}
|j\text{ }\gamma \rangle _\Pi  &\equiv &\tilde{R}(\gamma )|jj\rangle  
\nonumber \\
&=&\exp [-\frac \theta 2(\tilde{J}_{+}e^{-i\phi }-\tilde{J}_{-}e^{i\phi
})]|jj\rangle .
\end{eqnarray}
We call $|j$ $\gamma \rangle _\Pi $ the parity SU(2) coherent state, because
the parity operator plays a central role in its definition. This term
follows from that of the parity (harmonic oscillator) coherent state\cite
{Spi95}. The state $|j$ $\gamma \rangle _\Pi $ is different from the SU(2)
coherent state due to the nontrivial introduction of the su(2) parity
operator $\Pi $.

The antinormally ordered rotation operator is

\begin{equation}
\tilde{R}(\gamma )=\exp (\gamma \tilde{J}_{-})(1+|\gamma |^2)^{-\tilde{J}%
_z}\exp (-\gamma ^{*}\tilde{J}_{+}).
\end{equation}
Using the above equation, we obtain

\begin{eqnarray}
&&|j\text{ }\gamma \rangle _\Pi =\frac 1{\sqrt{2}}[e^{-i\frac \pi 4}|j\text{ 
}-i(-1)^{2j}\gamma \rangle   \nonumber \\
&&+e^{i\frac \pi 4}|j\text{ }i(-1)^{2j}\gamma \rangle ].
\end{eqnarray}
The SU(2) parity coherent state is a superposition of two SU(2) coherent
states with a phase difference $\pi $. It is similar to the parity harmonic
oscillator coherent states\cite{Spi95,Yur86}.

The general entangled SU(2) coherent state, analogous to the entangled
coherent state (2.5), is discussed in Appendix A. Here we introduce a
specific SU(2) coherent state by employing the su(2) parity operator (2.8).
Let us consider two independent su(2) Lie algebras. From these two algebras,
we can define two new algebras

\begin{eqnarray}
\lbrack \tilde{J}_{+}^n,\tilde{J}_{-}^l] &=&2\delta _{nl}\tilde{J}_z^n,\text{
}[\tilde{J}_z^n,\tilde{J}_{\pm }^l]=\pm \delta _{nl}\tilde{J}_{\pm }^n, 
\nonumber \\
\text{ }[\tilde{J}_{-}^n,\tilde{J}_{-}^l] &=&0,
\end{eqnarray}
where

\begin{eqnarray}
\tilde{J}_{+}^n &=&J_{+}^n\Pi ,\text{ }\tilde{J}_{-}^n=\Pi J_{-}^n\text{, }%
\tilde{J}_z^n\equiv J_z^n,  \nonumber \\
\Pi  &=&(-1)^{\text{ }{\cal M}_1+{\cal M}_2},\text{ }n,l=1,2.
\end{eqnarray}
It is easy to see that the parity operator $\Pi $ satisfies $\Pi ^2=1$ and $%
\Pi ^{\dagger }=\Pi $. These two new su(2) representation are dependent on
each other. The SU(2) coherent state of the two original su(2) algebra is

\begin{equation}
|j\text{ }\vec{\gamma}\rangle =R(\vec{\gamma})|jj\rangle _1\otimes
|jj\rangle _2,
\end{equation}
where $\vec{\gamma}\equiv (\gamma _1,\gamma _2),$ $|j$ $\vec{\gamma}\rangle
\equiv |j$ $\gamma _1\rangle _1\otimes |j$ $\gamma _2\rangle _2,$ and $R(%
\vec{\gamma})\equiv R_1(\gamma _1)R_2(\gamma _2).$ It is important that both
Hilbert spaces concerned in the entanglement are restricted to the same
irreducible representation $j$ for the entangled coherent states as we
introduce them to be well-defined for the  $j=1/2$ case. We consider
entangled qubit states; hence the restriction to the same $j$ is of
particular value as well. However the SU(2) coherent state for the two new
su(2) representations is obtained as

\begin{eqnarray}
|j\text{ }\vec{\gamma}\rangle _\Pi &=&\tilde{R}(\vec{\gamma})|jj\rangle
_1\otimes |jj\rangle _2  \nonumber \\
&=&\frac 1{\sqrt{2}}\left( e^{-i\frac \pi 4}|j\text{ }-i\vec{\gamma}\rangle
+e^{i\frac \pi 4}|j\text{ }i\vec{\gamma}\rangle \right)
\end{eqnarray}
The state $|j$ $\vec{\gamma}\rangle _\Pi $ is an entangled SU(2) coherent
state with a two-particle parity symmetry.

\subsection{Entangled SU(1,1) coherent states}

The generators of su(1,1) Lie algebras, $K_{\pm }$ and $K_z$, satisfy the
commutation relations

\begin{equation}
\lbrack K_{+},K_{-}]=-2K_z,\text{ }[K_z,K_{\pm }]=\pm K_{\pm }.
\end{equation}
By analogy to the su(2) case, we can define the su(1,1) parity operator as

\begin{equation}
\Pi =(-1)^{{\cal N}},\text{ }\Pi ^2=1,\text{ }\Pi ^{\dagger }=\Pi .
\end{equation}
where the `number' operator ${\cal N}$ is given by

\begin{equation}
{\cal N}=K_z-k,\text{ }{\cal N}|k\text{ }n\rangle =n|k\text{ }n\rangle .
\end{equation}
Here $|k$ $n\rangle $ $(n=0,1,2,...)$ is the complete orthonormal basis and $%
k\in \{1/2,1,3/2,2,...\}$ is the Bargmann index labeling the irreducible
representation[$k(k-1)$ is the value of the Casimir operator].

Using the su(1,1) parity operator we can introduce a new su(1,1) algebra
with generators

\begin{equation}
\tilde{K}_{+}=K_{+}\Pi ,\text{ }\tilde{K}_{-}=\Pi K_{-}\text{, }\tilde{K}%
_z\equiv K_z.
\end{equation}
There are two distinct SU(1,1) coherent states to consider.

\subsubsection{Entangled Perelomov SU(1,1) coherent states}

The Perelomov coherent state of the su(1,1) algebra is defined as\cite{Per86}

\begin{eqnarray}
|k\text{ }\eta \rangle _P &=&S(\xi )|k\text{ }0\rangle  \nonumber \\
&=&\exp (\xi K_{+}-\xi ^{*}K_{+})|k\text{ }0\rangle  \nonumber \\
&=&(1-|\eta |^2)^k\sum_{n=0}^\infty \sqrt{\frac{\Gamma (2k+n)}{\Gamma (2k)n!}%
}\eta ^n|k\text{ }n\rangle ,
\end{eqnarray}
where $\xi =r\exp (i\theta )$, $\eta =\exp (i\theta )\tanh r$, $\Gamma (x)$
is the Gamma function and $S(\xi )$ is the su(1,1) displacement operator.

We define a new SU(1,1) coherent state in association with the su(1,1)
parity operator

\begin{eqnarray}
|k\text{ }\eta \rangle _{P\Pi } &=&\tilde{S}(\xi )|k\text{ }0\rangle  
\nonumber \\
&=&\exp (\xi \tilde{K}_{+}-\xi ^{*}\tilde{K}_{-})|k\text{ }0\rangle ,
\end{eqnarray}
It is found that the Perelomov SU(1,1) coherent state is a nonlinear
coherent state with the nonlinear function $1/({\cal N}+2k)$\cite{Wang}.
Therefore, the parity Perelomov SU(1,1) coherent state is a nonlinear
coherent state with the nonlinear function$(-1)^{{\cal N}}/({\cal N}+2k).$

The normally-ordered su(1,1) displacement operator is
\begin{eqnarray}
\tilde{S}(\xi ) &=&\exp (\eta \tilde{K}_{+})(1-|\eta |^2)^{\tilde{K}_z}\exp
(-\eta ^{*}\tilde{K}_{-}),\text{ }  \nonumber \\
\eta &=&\frac \xi {|\xi |}\tanh (|\xi |).
\end{eqnarray}
Using the above equation, we obtain
\begin{eqnarray}
&&|k\text{ }\eta \rangle _{P\Pi }  \nonumber \\
&=&\frac 1{\sqrt{2}}\left( e^{i\frac \pi 4}|k\text{ }-i\eta \rangle _P+e^{-i%
\frac \pi 4}|k\text{ }i\eta \rangle _P\right) .
\end{eqnarray}
The parity Perelomov SU(1,1) coherent states are superpositions of two
Perelomov SU(1,1) coherent states.

The general entangled SU(1,1) coherent state is treated in Appendix A, but
here we consider the two-particle case. From the two su(1,1) algebras for
the two Hilbert space concerned with the entanglement, we define two new
su(1,1) algebras as
\begin{eqnarray}
\lbrack \tilde{K}_{+}^n,\tilde{K}_{-}^l] &=&-2\delta _{nl}\tilde{K}_z^n,%
\text{ }  \nonumber \\
\lbrack \tilde{K}_z^n,\tilde{K}_{\pm }^l] &=&\pm \delta _{nl}\tilde{K}_{\pm
}^l,\text{ }[\tilde{K}_{-}^n,\tilde{K}_{-}^l]=0.
\end{eqnarray}
where
\begin{eqnarray}
\tilde{K}_{+}^n &=&K_{+}^n\Pi ,\text{ }\tilde{K}_{-}^n=\Pi K_{-}^n\text{, }%
\tilde{K}_z^n\equiv K_z^n,  \nonumber \\
\Pi  &=&(-1)^{{\cal N}_1+{\cal N}_2},\text{ }n,l=1,2.
\end{eqnarray}
The su(1,1) parity operator $\Pi $ satisfies $\Pi ^2=1$ and $\Pi ^{\dagger
}=\Pi $.

The Perelomov SU(1,1) coherent state of the two new su(1,1) algebras is
obtained as
\begin{eqnarray}
|k\text{ }\vec{\eta}\rangle _{P\Pi } &=&\tilde{S}(\vec{\xi})|k\text{ }\vec{0}%
\rangle   \nonumber \\
&=&\frac 1{\sqrt{2}}\left( e^{i\frac \pi 4}|k\text{ }-i\vec{\eta}\rangle
_P+e^{-i\frac \pi 4}|k\text{ }i\vec{\eta}\rangle _P\right) 
\end{eqnarray}
Here $\tilde{S}(\vec{\xi})\equiv \tilde{S}_1(\xi _1)\tilde{S}_2(\xi _2).$
The state $|k$ $\vec{\eta}\rangle _{P\Pi }$ is the entangled Perelomov
SU(1,1) coherent state. Again each Hilbert space is restricted to the same
irrep $k,$ similar to the restriction for the su(2) case.

\subsubsection{Entangled Barut-Girardello SU(1,1) coherent states}

There is another coherent state of the su(1,1) algebra known as the
Barut-Girardello coherent state\cite{BGCS}. It is defined as the eigenstate
of the lowering operator $K_{-}$
\begin{equation}
K_{-}|k\text{ }\eta \rangle _{BG}=\eta |k\text{ }\eta \rangle _{BG},
\end{equation}
and it can be expressed as\cite{BGCS}
\begin{equation}
|k\text{ }\eta \rangle _{BG}=\sqrt{\frac{|\eta |^{2k-1}}{I_{2k-1}(2|\eta |)}}%
\sum_{n=0}^\infty \frac{\eta ^n}{\sqrt{n!\Gamma (n+2k)}}|k\text{ }n\rangle ,
\end{equation}
where $I_\nu (x)$ is the modified Bessel function of the first kind. The
Perelomov coherent state is defined with respect to the displacement
operator formalism, whereas the Barut-Girardello coherent state is defined
with respect to the ladder operator formalism. Thus, we define the parity
Barut-Girardello SU(1,1) coherent state as
\begin{equation}
\tilde{K}_{-}|k\text{ }\eta \rangle _{BG\Pi }=(-1)^{{\cal N}}K_{-}|k\text{ }%
\eta \rangle _{BG\Pi }=\eta |k\text{ }\eta \rangle _{BG\Pi }.
\end{equation}
The state $|k$ $\eta \rangle _{BG\Pi }$ is a nonlinear coherent state with
the nonlinear function $(-1)^{{\cal N}}$. From the general expression of a
SU(1,1) nonlinear coherent state\cite{Wang}, we obtain the expression of the
state $|k$ $\eta \rangle _{BG\Pi }$ as
\begin{equation}
|k\text{ }\eta \rangle _{BG\Pi }=\frac 1{\sqrt{2}}\left( e^{i\frac \pi 4}|k%
\text{ }-i\eta \rangle _{BG}+e^{-i\frac \pi 4}|k\text{ }i\eta \rangle
_{BG}\right) .
\end{equation}

Eigenstates of operators $\tilde{K}_{-}^n(n=1,2)$ are constructed as 
\begin{eqnarray}
|k\text{ }\vec{\eta}\rangle _{BG\Pi } &\sim &\exp \left( \frac{\eta _1}{%
{\cal N}_1+2k-1}\tilde{K}_{+}^1\right)  \\
&&\exp \left( \frac{\eta _2}{{\cal N}_2+2k-1}\tilde{K}_{+}^2\right) |k\text{ 
}\vec{0}\rangle   \nonumber \\
&=&\frac 1{\sqrt{2}}\left( e^{i\frac \pi 4}|k\text{ }-i\vec{\eta}\rangle
_{BG}+e^{-i\frac \pi 4}|k\text{ }i\vec{\eta}\rangle _{BG}\right) ,  \nonumber
\end{eqnarray}
which is the entangled Barut-Girardello SU(1,1) coherent states. Here we
have used the exponential form of the unnormalized Barut-Girardello coherent
state\cite{Wang}
\begin{equation}
|k\text{ }\eta \rangle _{BG}\sim \exp \left( \frac \eta {({\cal N}+2k-1)}%
K_{+}\right) |k\text{ }0\rangle .
\end{equation}
The generalization of (2.19), (2.30) and (2.35) to multiparticle entangled
SU(2) and SU(1,1) coherent states is treated in Appendix A.

\section{Nonlinear su(2) and su(1,1) coherent states}

\subsection{SU(2) case}

The su(2) parity operator $\Pi =(-1)^{{\cal M}}$ is a special case of the
unitary operator $U({\cal M})=\exp [-i\vartheta ({\cal M})])$. Here $%
\vartheta ({\cal M})$ is a nonsingular function of ${\cal M}.$ It is easy to
check that the operators

\begin{equation}
\bar{J}_{+}=J_{+}U({\cal M}),\text{ }\bar{J}_{-}=U^{\dagger }({\cal M})J_{-}%
\text{, }\bar{J}_z\equiv J_z
\end{equation}
satisfy the su(2) commutation relations (2.7). Then we can define the SU(2)
coherent state corresponding to this su(2) algebra as

\begin{eqnarray}
|\vartheta ;\text{ }j\text{ }\gamma \text{ }\rangle &\equiv &\bar{R}(\gamma
)|jj\rangle  \nonumber \\
&=&\exp \left[ -\frac 12\theta \left( \bar{J}_{+}e^{-i\phi }-\bar{J}%
_{-}e^{i\phi }\right) \right] |jj\rangle .
\end{eqnarray}

By use of (2.14), the state $|\vartheta ;$ $j$ $\gamma \rangle $ is obtained
as
\begin{eqnarray}
&&|\vartheta ;\text{ }j\text{ }\gamma \text{ }\rangle =\left( 1+|\gamma
|^2\right) ^{-j}\sum_{m=0}^{2j}{%
{2j \choose m}%
}^{1/2}\gamma ^m  \nonumber \\
&&\times \exp \left[ i\sum_{n=1}^m\vartheta (2j-n)\right] |j\,j-m\rangle ,
\end{eqnarray}
In the derivation of the above equation, we have used the relation

\begin{equation}
\lbrack f({\cal M})J_{-}]^m=(J_{-})^m\prod_{n=1}^mf({\cal M}-n)
\end{equation}

If we choose $\vartheta ({\cal M})=\pi {\cal M},$ equation (3.3) reduces to
(2.15) as we expected. We refer to the state $|\vartheta ;$ $j$ $\gamma $ $%
\rangle $ as a nonlinear SU(2) coherent state if $\vartheta ({\cal M})$ is a
nonlinear function of ${\cal M}$.

\subsection{SU(1,1) case}

By analogy to the su(2) case (3.1), we define

\begin{equation}
\bar{K}_{+}=K_{+}V({\cal N}),\text{ }\bar{K}_{-}=V^{\dagger }({\cal N})K_{-}%
\text{, }\bar{K}_z\equiv K_z.
\end{equation}
where $V({\cal N})=\exp [-i\varphi ({\cal N)}]$. These operators satisfy the
su(1,1) commutation relations (2.20). Then we can define the Perelomov
SU(1,1) coherent state corresponding to this su(1,1) algebra as

\begin{eqnarray}
|\varphi ;\text{ }k\text{ }\eta \rangle _P &=&\bar{S}(\xi )|k\text{ }0\rangle
\nonumber \\
&=&\exp (\xi \bar{K}_{+}-\xi ^{*}\bar{K}_{-})|k\text{ }0\rangle .
\end{eqnarray}

As is mentioned in the last section, the Perelomov SU(1,1) coherent state $%
|k $ $\eta \rangle _P$ is a nonlinear coherent state with the nonlinear
function $1/({\cal N}+2k).$ Therefore, the state $|\varphi ;$ $k$ $\eta
\rangle _P$ is a nonlinear coherent state with the nonlinear function $\exp
[i\varphi ({\cal N)}]/({\cal N}+2k).$ From the general expression for an
su(1,1) nonlinear coherent\cite{Wang}, we obtain

\begin{eqnarray}
|\varphi ;\text{ }k\text{ }\eta \rangle _P &=&(1-|\eta
|^2)^k\sum_{n=0}^\infty \sqrt{\frac{\Gamma (2k+n)}{\Gamma (2k)n!}}\eta ^n 
\nonumber \\
&&\times \exp \left[ -i\prod_{m=0}^{n-1}\varphi (m)\right] |k\text{ }%
n\rangle .
\end{eqnarray}
In the derivation of the above equation, we have used the relation

\begin{equation}
\lbrack K_{+}f({\cal N})]^m=(K_{+})^m\prod_{n=1}^mf({\cal N}+n-1).
\end{equation}

The nonlinear Barut-Girardello coherent state is defined as

\begin{eqnarray}
\bar{K}_{-}|\varphi ;\text{ }k\text{ }\eta \rangle _{BG} &=&\exp [i\varphi (%
{\cal N)}]K_{-}|\varphi ;\text{ }k\text{ }\eta \rangle _{BG}  \nonumber \\
&=&\eta |\varphi ;\text{ }k\text{ }\eta \rangle _{BG}.
\end{eqnarray}
The state $|\varphi ;$ $k$ $\eta \rangle _{BG}$ is also a nonlinear coherent
state with the nonlinear function $\exp [i\varphi ({\cal N)}].$ The
expansion of this state yields

\begin{eqnarray}
|\varphi ;\text{ }k\text{ }\eta \rangle _{BG} &=&\sqrt{\frac{|\eta |^{2k-1}}{%
I_{2k-1}(2|\eta |)}}\sum_{n=0}^\infty \frac 1{\sqrt{\Gamma (2k+n)n!}}\eta ^n
\nonumber \\
&&\times \exp [-i\prod_{m=0}^{n-1}\varphi (m)]|k\text{ }n\rangle .
\end{eqnarray}

If we choose $\varphi ({\cal N)=}\pi {\cal N},$ equations (3.7) and (3.10)
reduce to (2.27) and (2.34), respectively, as we expected.

\section{Representation of entangled su(2) and su(1,1) coherent states in
Fock space}

\subsection{Entangled binomial states}

It is well known that the operators

\begin{equation}
J_{+}=a^{\dagger }\sqrt{M-N}\text{, }J_{-}=\sqrt{M-N}a,\text{ }J_z=N-M/2
\end{equation}
generate the su(2) algebra via the Holstein-Primakoff representation\cite{HP}
in the spin $M/2$ representation. Here $N=a^{\dagger }a$ is the number
operator. The vacuum $|0\rangle $ is the lowest weight state

\begin{equation}
J_{-}|0\rangle =0,\text{ }J_z|0\rangle =-M/2|0\rangle .
\end{equation}

Then we define the SU(2) coherent state in the displacement operator form

\begin{eqnarray}
|M\text{ }\eta \rangle &=&\exp (\zeta J_{+}-\zeta ^{*}J_{-})|0\rangle 
\nonumber \\
&=&\exp (\zeta a^{\dagger }\sqrt{M-N}-\zeta ^{*}\sqrt{M-N}a)|0\rangle
\end{eqnarray}
where we choose $\zeta =\exp (i\theta )\arctan (|\eta |/\sqrt{1-|\eta |^2}),$
$\eta =|\eta |\exp (i\theta ).$ The normally ordered form of the
displacement operator is

\begin{eqnarray}
&&\exp (\zeta J_{+}-\zeta ^{*}J_{-}) \\
&&=\exp (\tilde{\eta}J_{+})(1+|\tilde{\eta}|^2)^{J_z}\exp (-\tilde{\eta}%
^{*}J_{-})  \nonumber
\end{eqnarray}
where $\tilde{\eta}=\zeta /|\zeta |\tan |\zeta |=\eta /\sqrt{1-|\eta |^2}.$
By use of (4.4), we obtain

\begin{equation}
|M\text{ }\eta \rangle =(1-|\eta |^2)^{M/2}\sum_{n=0}^M%
{M \choose n}%
^{1/2}\left( \frac \eta {\sqrt{1-|\eta |^2}}\right) ^n|n\rangle .
\end{equation}
This is the binomial state\cite
{Stoler85,Lee85,Dottoli87,Barranco94,Fan94,Joshi87}. Therefore, the binomial
state is a special type of SU(2) coherent state via the Holstein-Primakoff
representation. From (2.8) and (4.1) we see that the parity operator is $%
(-1)^N.$ By analogy to the discussion in Section II, we can obtain the
entangled binomial states of the type (2.19). Now we show how to generate
entangled binomial states in a particular Hamiltonian system.

The entangled coherent states can be
created using an ideal Kerr nonlinearity with three nonlinear media
elements\cite{San99}. We show that the entangled binomial state can also be generated in
this system. By an appropriate arrangement of the three nonlinear elements,
the effective Kerr transformation with two input fields, 1 and 2, is given by%
\cite{San99}

\begin{equation}
S_{12}(\chi )=\exp (-i\chi a_1^{\dagger }a_1a_2^{\dagger }a_2).
\end{equation}

We assume that the initial state is a product of binomial states $|M$ $\vec{%
\eta}\rangle =|M$ $\eta _1\rangle _1\otimes |M$ $\eta _2\rangle _2.$ For $%
\chi =\pi ,$ the resulting output state is

\begin{eqnarray}
&&\frac 12(|M\text{ }\eta _1\rangle _1+|M\text{ }-\eta _1\rangle _1)\otimes
|M\text{ }\eta _2\rangle _2  \nonumber \\
&&+(|M\text{ }\eta _1\rangle _1-|M\text{ }-\eta _1\rangle _1)\otimes |M\text{
}-\eta _2\rangle _2
\end{eqnarray}
This state contains both even and odd binomial states\cite{Barranco95}.

\subsection{Entangled negative binomial states}

The generators of the su(1,1) algebra via the Holstein-Primakoff realization
of the discrete irreducible representation with Bargmann index $M/2$ are

\begin{equation}
K_{+}=a^{\dagger }\sqrt{M+N}\text{, }K_{-}=\sqrt{M+N}a,\text{ }K_z=N+M/2,
\end{equation}
and the vacuum $|0\rangle $ is the lowest-weight state:

\begin{equation}
K_{-}|0\rangle =0,\text{ }K_z|0\rangle =M/2|0\rangle .
\end{equation}

The corresponding SU(1,1) coherent state is

\begin{equation}
|M\text{ }\eta \rangle ^{-}=\exp (\xi K_{+}-\xi ^{*}K_{-})|0\rangle .
\end{equation}
Here $\xi =\exp (i\theta )\arctan $h$|\eta |.$ Using (2.26), we obtain

\begin{equation}
|M\text{ }\eta \rangle ^{-}=(1-|\eta |^2)^{M/2}\sum_{n=0}^\infty {%
{M+n-1 \choose n}%
}^{1/2}\eta ^n|n\rangle .
\end{equation}
This is the negative binomial state\cite{Joshi89,Matsuo90,Joshi91,Agarwal92}%
. Therefore, the negative binomial state is a special type of SU(1,1)
Perelomov coherent state. The parity operator is $(-1)^N$ and the entangled
negative binomial states of the type (2.30) can be obtained. We do not write
these entangled states explicitly here.

Under the transformation $S_{12}(\pi )$, the input negative binomial state $%
|M$ $\vec{\eta}\rangle ^{-}$ will be transformed into the entangled negative
binomial states

\begin{eqnarray}
&&\frac 12(|M\text{ }\eta _1\rangle _1^{-}+|M\text{ }-\eta _1\rangle
_1^{-})\otimes |M\text{ }\eta _2\rangle _2^{-}  \nonumber \\
&&+(|M\text{ }\eta _1\rangle _1^{-}-|M\text{ }-\eta _1\rangle _1^{-})\otimes
|M\text{ }-\eta _2\rangle _2^{-}.
\end{eqnarray}
This state contains both even and odd negative binomial states\cite
{Joshi97,Wang2}

\subsection{Contraction of su(2) and su(1,1) algebra}

Let us note the fact that the binomial distribution tends to the Poisson
distribution in a certain limit. Let $M\rightarrow \infty ,|\eta
|\rightarrow 0$ in such a way that the product $|\eta |^2M=|\alpha |^2$ is
fixed. In this limit, the binomial distribution of the binomial state tends
to the Poisson distribution $\exp (-|\alpha |^2)|\alpha |^{2n}/n!,$ and the
binomial state tends to the ordinary coherent state 
\begin{equation}
|M\text{ }\eta \rangle \rightarrow \exp (-|\alpha |^2/2)\sum_{n=0}^\infty 
\frac{\alpha ^n}{\sqrt{n!}}|n\rangle .
\end{equation}
Here we have used the relation

\begin{equation}
(1-|\eta |^2)^M\rightarrow \exp (-|\alpha |^2).
\end{equation}

This limit can also be visualized as a contraction of the su(2) algebra into
the Heisenberg-Weyl algebra\cite{Fu} 
\begin{equation}
|\eta |J_{+}\rightarrow |\alpha |a^{\dagger },|\eta |J_{-}\rightarrow
|\alpha |a.
\end{equation}
Thus, (4.5) tends to the coherent state 
\begin{equation}
|M\text{ }\eta \rangle \rightarrow \exp [\alpha a^{\dagger }-\alpha
^{*}a]|0\rangle .
\end{equation}

In this limit the entangled binomial state (4.7) reduces to the entangled
coherent state\cite{San99}

\begin{eqnarray}
&&\frac 12[(|\alpha _1\rangle _1+|-\alpha _1\rangle _1)\otimes |\alpha
_2\rangle _2 \\
&&+(|\alpha _1\rangle _1-|-\alpha _1\rangle _1)\otimes |-\alpha _2\rangle _2
\nonumber
\end{eqnarray}
which can be employed as qubits in quantum computation\cite
{Munro99,Oliveira00}.

In the same limit described above, the negative binomial states reduce to
coherent states, the su(1,1) algebra contracts into the Heisenberg-Weyl
algebra\cite{Fu}, and the entangled negative binomial states (4.12) reduce
to the entangled harmonic oscillator coherent states (4.17).

\subsection{Entangled squeezed states}

The amplitude-squared su(1,1) algebra is realized by

\begin{equation}
K_{+}=\frac 12a^{+2},K_{-}=\frac 12a^2,K_z=\frac 12\left( N+\frac 12\right) .
\end{equation}
The representation on the usual Fock space is completely reducible and
decomposes into a direct sum of the even Fock space ($S_0$) and odd Fock
space ($S_1$),

\begin{equation}
S_i=\text{span}\{|n\rangle _i\equiv |2n+i\rangle |n=0,1,2,...\},\text{ }%
i=0,1.
\end{equation}

Representations on $S_i$ can be written as

\begin{eqnarray}
K_{+}|n\rangle _i &=&\sqrt{(n+1)(n+i+1/2)}|n+1\rangle _i, \\
K_{-}|n\rangle _i &=&\sqrt{(n)(n+i-1/2)}|n-1\rangle _i,  \nonumber \\
K_0|n\rangle _i &=&(n+i/2+1/4)|n\rangle _i.  \nonumber
\end{eqnarray}
The Bargmann index $k=1/4$ ($3/4$) for even (odd) Fock space. We see that
the Perelomov SU(1,1) coherent states in even/odd Fock space are squeezed
vacuum states and squeezed first Fock states

\begin{eqnarray}
|\eta \rangle _V &=&\exp \left( \frac \xi 2a^{+2}-\frac{\xi ^{*}}2a^2\right)
|0\rangle , \\
|\eta \rangle _F &=&\exp \left( \frac \xi 2a^{+2}-\frac{\xi ^{*}}2a^2\right)
|1\rangle ,  \nonumber
\end{eqnarray}
respectively. From (2.21) and (4.18), the corresponding parity operator is $%
(-1)^{N/2}$ in even Fock space and $(-1)^{(N-1)/2}$ in odd Fock space. We
consider two modes $a_1$ and $a_2.$ Then from (2.30) and through the
amplitude-squared su(1,1) realization, we obtain the entangled squeezed
vacuum states and entangled squeezed first Fock states as

\begin{eqnarray}
&&|\vec{\eta}\rangle _V=\frac 1{\sqrt{2}}\left( e^{i\frac \pi 4}|-i\vec{\eta}%
\rangle _V+e^{-i\frac \pi 4}|i\vec{\eta}\rangle _V\right) ,  \nonumber \\
&&|\vec{\eta}\rangle _F=\frac 1{\sqrt{2}}\left( e^{i\frac \pi 4}|-i\vec{\eta}%
\rangle _F+e^{-i\frac \pi 4}|i\vec{\eta}\rangle _F\right) .
\end{eqnarray}
The state (4.22) is a special case of the entangled squeezed coherent state%
\cite{San95} for zero amplitude and reduces to the superposition of two
squeezed vacuum states\cite{Sanders89}.

In Fock space, we have obtained entangled binomial states, entangled
negative binomial states and entangled squeezed states. They are special
cases of entangled SU(2) coherent states or entangled SU(1,1) coherent
states.

\section{Generation of the entangled coherent states}

A superposition of two distinct su(2) coherent states can be generated by
the Hamiltonian system for the nonlinear rotator\cite{San89,Kiti93}

\begin{equation}
H_j=\omega J_z+\frac \lambda {2j}J_z^2\text{ (}\hbar =1\text{)}
\end{equation}
where $\omega $ is the linear precession frequency and $\lambda $ is a
positive constant. Let the initial state be the SU(2) coherent state $%
|j,\gamma \rangle .$ At time $t_j=\pi j/\lambda $, the coherent state has
evolved into the superposition state
\begin{eqnarray}
&&\exp (-iH_jt_j)|j\text{ }\gamma \rangle  \nonumber \\
&=&\frac 1{\sqrt{2}}\left( e^{-i\frac \pi 4}|j\text{ }\bar{\gamma}\rangle
+e^{i\frac \pi 4}(-1)^j|j\text{ }-\bar{\gamma}\rangle \right) .
\end{eqnarray}
Here $\bar{\gamma}=\exp (i\omega \pi j/\lambda )\gamma .$ A superposition of
two coherent states, separated by a phase $\pi ,$ has been generated from
the initial SU(2) coherent state. The phase difference of the coefficients
in (5.2) depends on whether $j$ is odd or even. Let $\bar{\gamma}\rightarrow
-i(-1)^{2j}\gamma $ and $j$ be even; the above state is just the parity
SU(2) coherent state $|j$ $\gamma \rangle _\Pi .$

Gerry\cite{Gerry87} and Bu$\check{z}$ek\cite{Buzek89} have studied the
dynamics of the nonlinear oscillator with the Hamiltonian $H$ written as
\begin{equation}
H=\omega a^{\dagger }a+\frac \lambda 2a^{\dagger 2}a^2=\omega K_z+\lambda
K_{+}K_{-}
\end{equation}
up to constant terms. Here $K_z$ and $K_{\pm }$ are generators of the
amplitude-squared su(1,1) algebra (4.18). At time $t=\pi /2\lambda $, the
initial coherent state has evolved into the superposition state
\begin{eqnarray}
&&\frac 1{\sqrt{2}}\left[ e^{-i\frac \pi 4}|k\text{ }\bar{\eta}\rangle
_P+e^{i\frac \pi 4}|k\text{ }-\bar{\eta}\rangle _P\right] , \\
&&\frac 1{\sqrt{2}}\left[ e^{-i\frac \pi 4}|k\text{ }\bar{\eta}\rangle
_{BG}+e^{i\frac \pi 4}|k\text{ }-\bar{\eta}\rangle _{BG}\right] ,
\end{eqnarray}
where $\bar{\eta}=\exp \{-i\pi [\omega +(2k-1)\lambda ]/(2\lambda )\}.$ Here
we choose the initial state to be the SU(1,1) coherent state $|k$ $\eta
\rangle _P$ and $|k$ $\eta \rangle _{BG}$, respectively. Let $\bar{\eta}%
\rightarrow i\eta ;$ the above superposition states are then just the parity
coherent states $|k\, \eta \rangle _{P\Pi }$ and $|k$ $\eta \rangle _{BG\Pi
} $.

Now we consider how to generate the entangled SU(2) and SU(1,1) coherent
states. By analogy with the generation of entangled coherent states of the
harmonic oscillator\cite{Spi95}, the two Hamiltonians are given by
\begin{eqnarray}
H_2 &=&\chi _1{\cal M}_1^2+\chi _2{\cal M}_2^2+\chi _3{\cal M}_1{\cal M}_2,
\\
H_{11} &=&\lambda _1{\cal N}_1^2+\lambda _2{\cal N}_2^2+\lambda _3{\cal N}_1%
{\cal N}_2.
\end{eqnarray}
We assume that the initial state of the Hamiltonian system $H_2$ is $|j$ $-i%
\vec{\gamma}\rangle $ and the time $t=\pi /(2\chi _1),\chi _2=\chi _1,\chi
_3=2\chi _1$. Then the initial state evolves into the entangled SU(2)
coherent state $|j$ $\vec{\gamma}\rangle _\Pi $ $(j$ is an integer$).$
Similarly, we assume that the initial state of the Hamiltonian system $H_{11}
$ is $|k$ $i\vec{\eta}\rangle _P$ and the time $t=\pi /(2\lambda _1),$ $%
\lambda _2=\lambda _1,$ $\lambda _3=2\lambda _1$. The initial state evolves
into the entangled SU(1,1) coherent state $|k$ $\vec{\eta}\rangle _{P\Pi 
\text{.}}$ If we choose the initial state as an SU(1,1) coherent state $|k$ $%
i\vec{\eta}\rangle _{BG},$ the resulting state will be the entangled SU(1,1)
coherent states $|k$ $\vec{\eta}\rangle _{BG\Pi \text{.}}$

It is interesting if we choose the parameters $\chi _1=\chi _2=0,t=\pi /\chi
_3.$ The initial state $|j\, \vec{\gamma}\rangle $ will evolve into the following entangled SU(2)
coherent state

\begin{eqnarray}
&&\frac 12\{[|j\text{ }\gamma _1\rangle _1+(-1)^{2j}|j\text{ }-\gamma
_1\rangle _1]\otimes |j\text{ }\gamma _2\rangle _2+ \\
&&[(-1)^{2j}|j\text{ }\gamma _1\rangle _1-|j\text{ }-\gamma _1\rangle
_1]\otimes |j\text{ }-\gamma _2\rangle _2\}  \nonumber \\
&&=\frac 12\{|j\text{ }\gamma _1\rangle _1\otimes [|j\text{ }\gamma
_2\rangle _2+(-1)^{2j}|j\text{ -}\gamma _2\rangle _2]+  \nonumber \\
&&|j\text{ }-\gamma _1\rangle _1\otimes [(-1)^{2j}|j\text{ }\gamma _2\rangle
_2-|j\text{ }-\gamma _2\rangle _2]\}.  \nonumber
\end{eqnarray}
The entangled coherent state (5.8)  includes the SU(2) `cat' states\cite
{San89,Gerry97}. We assume that the initial state of the Hamiltonian system $%
H_{11}$ is $|k$ $\vec{\eta}\rangle .$ Here the subscripts $P$ and $BG$ have
been omitted since the discussions are the same for the Perelomov coherent
states and the Barut-Girardello coherent states. At time $t=\pi /\lambda _3$
and $\lambda _1=\lambda _2=0,$ Resultant state is

\begin{eqnarray}
&&\frac 12[(|k\text{ }\eta _1\rangle _1+|k\text{ }-\eta _1\rangle _1)\otimes
|k\text{ }\eta _2\rangle _2+ \\
&&(|k\text{ }\eta _1\rangle _1-|k\text{ }-\eta _1\rangle _1)\otimes |k\text{ 
}-\eta _2\rangle _2],  \nonumber
\end{eqnarray}
which includes the SU(1,1) `cat' states\cite{Gerry97}.

\section{Degree of entanglement}

\subsection{Bell inequality}

A standard example of entanglement of two-particle nonorthogonal states is
given by\cite{Man95}

\begin{equation}
|\Psi \rangle =\mu |\alpha \rangle _1\otimes |\beta \rangle _2+\nu |\gamma
\rangle _1\otimes |\delta \rangle _2,
\end{equation}
where $|\alpha \rangle _1$ and $|\gamma \rangle _1$ are nonorthogonal states
of system 1 and similarly for $|\beta \rangle _2$ and $|\delta \rangle _2$
of system 2. We also assume that the states $|\alpha \rangle _1$ and $%
|\gamma \rangle _1$ are linearly independent as are $|\beta \rangle _2$ and $%
|\delta \rangle _2$. The entangled SU(2) and SU(1,1) coherent states
obtained in the previous section are special cases of these entangled
nonorthogonal states $|\Psi \rangle $. The state $|\Psi \rangle $ is a pure
state whose density matrix is $\rho _{12}=|\Psi \rangle \langle \Psi |.$
Then the reduced density matrices $\rho _1=$Tr$_2(\rho _{12})$ and $\rho _2=$%
Tr$_1(\rho _{12})$ for each subsystem can be obtained, and the two
eigenvalues of $\rho _1$ are given by\cite{Man95}

\begin{eqnarray}
\lambda _{\pm } &=&\frac 12\pm \frac 12\sqrt{1-4|\mu \nu {\cal A}_1{\cal A}%
_2|^2},  \nonumber \\
{\cal A}_1 &=&\sqrt{1-|_1\langle \alpha |\gamma \rangle _1|^2},\text{ }{\cal %
A}_2=\sqrt{1-|_2\langle \delta |\beta \rangle _2|^2}.
\end{eqnarray}
The two eigenvalues of $\rho _{2\text{ }}$are identical to those of $\rho _{1%
\text{.}}$ The corresponding eigenvectors of $\rho _1$ are $|\pm \rangle _1,$
and the corresponding orthonormal eigenvectors of $\rho _2$ are denoted by $%
|\pm \rangle _2.$ The general theory of the Schmidt decomposition\cite
{Everett57,Barnett92,Knight93} implies that the state $|\Psi \rangle $ can
be expressed in the Schmidt form

\begin{equation}
|\Psi \rangle =c_{-}|-\rangle _1|-\rangle _2+c_{+}|+\rangle _1|+\rangle _2,
\end{equation}
where $|c_{\pm }|^2=\lambda _{\pm },$ $|c_{+}|^2+|c_{-}|^2=1.$ The
two-system entangled state(6.3) violates a Bell inequality. More
specifically, we choose Hermitian operator $\hat{\Theta}_i$ for each
subsystem such that the eigenvalues are $\pm 1.$ The general form for such
an operator is

\begin{eqnarray}
\hat{\Theta}_i(\lambda _i,\varphi _i) &=&\cos \lambda _i(|+\rangle _i\langle
+|-|-\rangle _i\langle -|)  \nonumber \\
&&+\sin \lambda _i(e^{i\varphi _i}|+\rangle _i\langle -|+e^{-i\varphi
_i}|-\rangle _i\langle +|).
\end{eqnarray}

The Bell operator is defined as\cite{Braunstein92}

\begin{equation}
\hat{B}=\hat{\Theta}_1\hat{\Theta}_2+\hat{\Theta}_1\hat{\Theta}_2^{\prime }+%
\hat{\Theta}_1^{\prime }\hat{\Theta}_2-\hat{\Theta}_1^{\prime }\hat{\Theta}%
_2^{\prime },
\end{equation}
\newline
for $\hat{\Theta}_i\equiv \hat{\Theta}_i(\lambda _i,\varphi _i),$ $\hat{%
\Theta}_i^{\prime }\equiv \hat{\Theta}_i(\lambda _i^{\prime },\varphi
_i^{\prime }).$

For the choices

\begin{eqnarray}
\lambda _1 &=&0,\text{ }\lambda _1^{\prime }=\pi /2, \\
\lambda _2 &=&-\text{ }\lambda _2^{\prime }=\cos
^{-1}[1+|2c_{+}c_{-}|^2]^{-1/2},  \nonumber \\
\varphi _1+\varphi _2 &=&\varphi _1^{\prime }+\varphi _2^{\prime }=\varphi
_{+}-\varphi _{-},  \nonumber
\end{eqnarray}
where $\varphi _{\pm }$ are the phases of $c_{\pm }$ in (6.3), the
expectation value of the Bell operator for the state $|\Psi \rangle $ $(6.1)$
is\cite{Man95}

\begin{equation}
B=\langle \Psi |\hat{B}|\Psi \rangle =2\sqrt{1+4\lambda _{+}\lambda _{-}}>2.
\end{equation}
The degree of violation depends on the values of $\lambda _{\pm },$ but a
violation always occurs provided that the state is entangled.

Several entangled coherent states are obtained in the previous sections.
Here we only consider two examples. The first example is the entangled SU(2)
coherent state $|j$ $\vec{\gamma}\rangle _\Pi $ (2.19). For simplicity $%
\gamma _0=\gamma _1=\gamma _2$ is introduced. The eigenvalues of the reduced
density matrices $\rho _1$ and $\rho _2$ are

\begin{equation}
\lambda _{\pm }=\frac 12\pm \frac 12\sqrt{1-\left[ 1-\left( \frac{1-|\gamma
_0|^2}{1+|\gamma _0|^2}\right) ^{4j}\right] ^2}.
\end{equation}
Then the expectation value of the Bell operator for the entangled coherent
state $|j$ $\vec{\gamma}\rangle _\Pi $ is given by

\begin{equation}
B=2\sqrt{1+\left[ 1-\left( \frac{1-|\gamma _0|^2}{1+|\gamma _0|^2}\right)
^{4j}\right]^2}.
\end{equation}

The second example is the entangled Perelomov SU(1,1) coherent state $|k\,%
\vec{\eta}\rangle _{P\Pi }$. The corresponding eigenvalues and expectation
value of the Bell operator are obtained as 
\begin{eqnarray}
\lambda _{\pm } &=&\frac 12\pm \frac 12\sqrt{1-\left[ 1-\left( \frac{1-|\eta
_0|^2}{1+|\eta _0|^2}\right) ^{4k}\right] ^2}, \\
B &=&2\sqrt{1+\left[ 1-\left( \frac{1-|\eta _0|^2}{1+|\eta _0|^2}\right)
^{4k}\right] ^2}.
\end{eqnarray}
Here we have chosen $\eta _0=\eta _1=\eta _2.$

\subsection{Entropy}

The entropy $S$ of a quantum state described by density operator $\rho $ is
defined by\cite{Araki70,Wehrl78,Barnett89}

\begin{equation}
S=-k_B\text{Tr(}\rho \ln \rho ),
\end{equation}
where $k_B$ is the Boltzmann constant. The entropy defined above is zero for
a pure state and positive for a mixed state. If we consider two systems, the
entropy the system 1(2) is determined via the reduced density operator

\begin{equation}
S_{1(2)}=-k_B\text{Tr}_{1(2)}\text{(}\rho _{1(2)}\ln \rho _{1(2)}).
\end{equation}

In 1970 Araki and Lieb proved the following inequality\cite{Araki70}:

\begin{equation}
|S_1-S_2|\leq S\leq S_1+S_2.
\end{equation}
One consequence of this inequality is that, if the total system is in a pure
state, $S_1=S_2.$ From an information theory point of view, the entropy can
be regarded as the amount of uncertainty contained within the density
operator. We can use the index of correlation $I_c$ as the amount of
information lost in the tracing procedure\cite{Barnett89}

\begin{equation}
I_c=S_1+S_2-S.
\end{equation}

For the pure state $|\Psi \rangle ,$ the index of correlation is obtained as

\begin{equation}
I_c=2S_1=-2k_B(\lambda _{+}\ln \lambda _{+}+\lambda _{-}\ln \lambda _{-}).
\end{equation}
If $I_c=0,$ the two subsystems are in a pure state and disentangled. The
combination of (6.8-6.11) and (6.16) gives the index of correlation of
entangled SU(2) and SU(1,1) coherent states.

\subsection{Discussion}

We now discuss two limit cases for certain parameters. When %
\mbox{$\vert$}$\gamma _0|(|\eta _0|)\rightarrow 0$ or $\infty ,$ $B=2,$
i.e., the entangled coherent states become product states and disentangled.
When \mbox{$\vert$}$\gamma _0|(|\eta _0|)=1,$ $B=2\sqrt{2},$ and the
entangled coherent states are maximally entangled states. For this case, the
two states $|j$ $\gamma _0\rangle $ and $|j$ $-\gamma _0\rangle $ are
orthogonal and the states $|k$ $\eta _0\rangle $ and $|k$ $-\eta _0\rangle $
become orthogonal.

When \mbox{$\vert$}$\gamma _0|(|\eta _0|)\rightarrow 0$ or $\infty ,$ $%
\lambda _{+}=1$ and $\lambda _{-}=0,$ then the index of correlation is zero
as we expect. The entangled states become product states. When \mbox{$\vert$}%
$\gamma _0|(|\eta _0|)=1,$ $\lambda _{\pm }=1/2,$ then the index of
correlation is $2k_B\ln 2,$ which results from the orthogonality of the
states $|j$ $\pm \gamma _0\rangle (|\gamma _0|=1)$ or $|k$ $\pm \eta
_0\rangle (|\eta _0|=1).$ An index of correlation is maximum for $I_c=2k_B
\ln 2$ \cite{Barnett89}.

The maximum entanglement of SU(2) coherent states $|j$ $\vec{\gamma}\rangle
_\Pi $ for $\vec{\gamma}=(\gamma _0,\gamma _0)$ and $|\gamma _0|=1$ can be
understood as follows. From the expression for $\gamma =\exp (i\phi )\tan (\theta /2),$ we can
see that $|\gamma |=1$ corresponds to the set of all points on the equator ($\theta=\pi/2$)
 of the Poincar$\acute{e}$ sphere, which is used to represent the pure states
of an SU(2) system. The states corresponding to $|j$ $\gamma _0=1\rangle $
and  $|j\, \gamma _0=-1\rangle $ are antipodal; that is, one state is
represented by a point at the intersection of the equator and the
longitudinal line $\phi =0,$ and the other state is represented by the
point at the intersection of the equator and the longitudinal line $\phi
=\pi .$ We can thus employ the notation of Eq. (1.1), but with an
appropriate SU(2) rotation (and allow for arbitrary $j$) by replacing $|j$ $%
\gamma _0=1\rangle $ by $|1\rangle $ and $|j$ $\gamma _0=-1\rangle $ by $%
|0\rangle .$ For $j=1/2,$ we recover the qubit case (1.1). The entanglement
state for $j=1/2$ is just $2^{-1/2}\left[ \exp (i\pi /4)|11\rangle +\exp
(-i\pi /4)|00\rangle \right] .$ This state is a form of Bell state (1.3) and
the reason for $B$ and $I_c$ being maximal is evident. A similar analysis
can be employed for the entangled SU(1,1) coherent states represented by
points on the Lobachevsky plane.

\section{Conclusions}

We have introduced entangled SU(2) and SU(1,1) coherent states. The general
form for these entangled coherent states, which also incorporates entangled
harmonic oscillator coherent states in the formalism, is given in the Appendix, but
the main concern here is with two-particle coherent states. The two-particle
coherent states present a diverse range of interesting phenomena, as we have
shown. Aside from the mathematical elegance of entangled SU(2) and SU(1,1)
coherent states, we have also applied these states to current topics of
research, namely quantum information (specifically qubits) and nonclassical
states of light (specifically squeezed vacuum states).

We have explored several aspects of coherent states. Aside from defining
these states, we have employed the entangled binomial states to establish a
rigorous contraction from entangled SU(2) coherent states to entangled
harmonic oscillator coherent states. Two measures of entanglement, the Bell
operator approach and the index of correlation, have been used to quantify
the degree of entanglement. As the entanglement is generally between
non-orthogonal states, the degree of entanglement ranges from no
entanglement (product state) to being maximally entangled for various
parametric choices.

The generation of entangled coherent states have been treated here by a
Hamiltonian evolution which is a generalization of $J_z^2$ and $K_z^2$
nonlinear evolution for multiparticle systems. However, such evolutions are
extremely sensitive to environmental-induced decoherence. Other methods for
generating entangled coherent states could be considered, but the nonlinear
evolution considered here illustrated one possible approach to producing
these entangled states.

These results can be generalized in various ways which would be of interest.
One generalization is to entangled generalized coherent states, with
generalized coherent states of the type treated by Perelomov\cite{Per86}.
Another intriguing generalization is to entangled SU(2) and SU(1,1) coherent
states for the respective Hilbert spaces, not restricted to the same 
irreducible representations. For example one could consider entanglement
between spin-1/2 states (qubits) and spin-1 states (qutrits). Or one could
consider entanglement of SU(1,1) coherent states with $k=1/4$ (single mode
squeezed vacuum states) and SU(1,1) coherent states with $k=3/4$ (two-mode
squeezed vacuum states). These ideas are  topics for future research.

\appendix

\section{Most General form of Entangled SU(2) AND SU(1,1) coherent states}

The general form of entangled SU(2) and SU(1,1) coherent states can be
written as

\begin{equation}
\int d\mu (\vec{\xi})f(\vec{\xi})|l\text{ }\vec{\xi}\rangle ,
\end{equation}
where $l=j$ and $\vec{\xi}=\vec{\gamma}$ for SU(2) coherent states, and $l=k$
and $\vec{\xi}=\vec{\eta}$ for either Perelomov or Barut-Girardello SU(1,1)
coherent states. We can also obtain the entangled coherent state (2.5) via
the same expression by replacing $\vec{\xi}$ by $\vec{\alpha}$ and ignoring
the irrep index $l.$ The measure $d\mu (\vec{\xi})$ is, for each case, 
\begin{eqnarray}
d\mu (\vec{\gamma}) &=&\prod_n\frac{2j+1}\pi \frac{d^2\vec{\gamma}_n}{%
(1+|\gamma _n|^2)^2},  \nonumber \\
d\mu _P(\vec{\eta}) &=&\prod_n\frac{2k-1}\pi \frac{d^2\vec{\eta}_n}{(1-|\eta
_n|^2)^2},  \nonumber \\
d\mu _{BG}(\vec{\eta}) &=&\prod_n\frac 2\pi I_{2k-1}^2(2|\eta _n|)\text{ }d^2%
\vec{\eta}_n,  \nonumber \\
d\mu (\vec{\alpha}) &=&\prod_n\frac{d^2\vec{\alpha}}\pi
\end{eqnarray}
for SU(2) coherent states, Perelomov SU(1,1) coherent states and
Barut-Girardello coherent states, and harmonic oscillator coherent states,
respectively.

If we choose $l=j$ and 
\begin{equation}
f(\vec{\xi})=\frac 1{\sqrt{2}}\left[ \exp \left( -i\frac \pi 4\right) \delta
(\vec{\xi}+i\vec{\gamma})+\exp \left( i\frac \pi 4\right) \delta (\vec{\xi}-i%
\vec{\gamma})\right] ,
\end{equation}
the general state (A1) will reduce to the entangled SU(2) coherent state
(2.19). All examples of coherent states considered in this paper can be
constructed accordingly.

As an interesting example, we consider the multiple qubit entanglement. Shor 
\cite{Shor} introduced the quantum Fourier transforms in order to apply
quantum computation to factorize a number $a$, with $0\leq a\leq q.$ This
number $a$ can be expressed in qubits as

\begin{equation}
|a\rangle \equiv |\vec{\varepsilon}\rangle =|\varepsilon _{1,}\text{ }%
\varepsilon _{2,}\text{ }\cdot \cdot \cdot ,\varepsilon _N\rangle ,\text{ }%
\varepsilon _i\in \{0,1\},
\end{equation}
\newline
where $\vec{\varepsilon}$ is the binary notation of $a.$ Here $N$ is the
lowest integer greater than or equal to $\log _2q.$ The state $|a\rangle $
is a product multiparticle SU(2) coherent state for $j=1/2.$ The quantum
Fourier transform of the state $|a\rangle $ is

\begin{equation}
q^{-1/2}\sum_{c=0}^{q-1}\exp (i2\pi ac/q)|c\rangle .
\end{equation}
The state $|c\rangle $ is also a product state for all $c$ such that 
$0 \leq c\leq q-1.$ We can express the transformed state as the entangled
SU(2) coherent state (A1) such that

\begin{eqnarray}
&&q^{-1/2}\sum_{\vec{\varepsilon}_i\in \{0,1\}}\exp (i2\pi ac/q)|\vec{%
\varepsilon}\rangle  \nonumber \\
&=&q^{-1/2}\sum_{\vec{\gamma}_i\in \{0,\infty \}}\exp (i2\pi ac/q)|j\text{ }%
\vec{\gamma}\rangle
\end{eqnarray}
where $c\equiv \vec{\varepsilon}.$ The above state can also be written in
the form of (A1) with
\begin{equation}
f(\vec{\xi})=q^{-1/2}\sum_{\vec{\gamma}_i\in \{0,\infty \}}\exp (i2\pi
ac/q)\delta (\vec{\xi}-\vec{\gamma}_i).
\end{equation}
Hence the Fourier transform state (A5) can be treated within the formalism of
entangled SU(2) coherent states.


\end{document}